\documentstyle[preprint,aps]{revtex}
\tightenlines

\begin{document}

\draft
\title
{Hausdorff dimension, fractional spin particles and 
Chern-Simons effective potential}

\author
{Wellington da Cruz\footnote{E-mail: wdacruz@fisica.uel.br}}

\address
{Departamento de F\'{\i}sica,\\
 Universidade Estadual de Londrina, Caixa Postal 6001,\\
Cep 86051-970 Londrina, PR, Brazil\\}
 
\date{\today}

\maketitle

\begin{abstract}
We obtain for any spin, $s$, the Hausdorff dimension,
 $h_{i}$, 
for fractional spin particles and we discuss the
 connection between 
this number, $h_{i}$, and the Chern-Simons potential.
 We also define 
the topological invariants, ${\cal W}_{s}$, in terms
 of the statistics 
of these particles.

\end{abstract}

\pacs{PACS numbers: 11.90+t\\
Keywords: Hausdorff dimension; Fractional spin particles;
 Topological invariants;
Chern-Simons potential }
%\narrowtext

%\

In this letter, for any spin, $s$, we obtain 
 the Hausdorff
dimension for fractional spin particles. We generalize
 a result
 obtained in\cite{R1}. Below, we give some examples 
 ( $i$ means a specific interval ):
 
\begin{eqnarray}
&&h_{1}=2-2s,\;\;\;\; 0\leq s\leq 0.5;\;\;\;\;\; h_{2}=2s,\;\;\;
\;\;\;\;\; 0.5\leq s \leq 1;\;\nonumber\\
&&h_{3}=4-2s,\;\;\;\; 1 \leq s \leq 1.5;\;\;\;\;\;
h_{4}=2s-2,\;\; 1.5\leq s \leq 2;\;\nonumber\\
&&h_{5}=6-2s,\;\;\;\; 2 \leq s \leq 2.5;\;\;\;\;\;
h_{6}=2s-4,\;\; 2.5\leq s \leq 3;\;\\
&&h_{7}=8-2s,\;\;\;\; 3\leq s \leq 3.5;\;\;\;\;\;
h_{8}=2s-6,\;\; 3.5\leq s \leq 4;\;\nonumber\\
&&h_{9}=10-2s,\;\;4\leq s\leq 4.5;\;\;\;
h_{10}=2s-8,\;\;\; 4.5\leq s\leq 5;\nonumber\\ 
&&etc.\nonumber
\end{eqnarray}
 
\noindent Each $h_{i}$ {\it per se}, within in its
 intervals
of definition, satisfies all properties of the
 Hausdorff dimension, such as, 
open sets, smooth sets, monotonicity, countable
 stability and countable 
sets\cite{R2}. On the other hand, we know that
 a charge-flux system
posseses a spin $s$
   
\begin{equation}
s=-\frac{e\Phi}{hc},
\end{equation}

\noindent where for this formula only, $h$ denotes Planck's
constant, $c$ is the velocity 
of light, $\Phi$ is the magnetic flux and $e$ 
is the electric charge; and the anyonic statistics is 
$\nu=2s$, with $s$ real\cite{R3}. 
Thus, all formulas for $h_{i}$ in each interval $s$ 
can be put in terms of the statistics. Therefore, 
we can define the topological 
invariants, ${\cal W}_{s}$, for an anyonic system 
with distinct spins in the same way that we did in\cite{R1}. 
Then, we see that ${\cal W}_{s}={\cal W}$ always, independent 
of the intervals of definition of $s$, for example, in the 
first interval,

\begin{equation}
{\cal W}=h+2s-2p=h+\nu-2p,
\end{equation}

\noindent where $p$ represents the number of fractional 
spin particles with statistics $\nu$ and Hausdorff 
dimension $h$. Another interesting point is that we have 
the same values of $h$ for particles with distinct spins. 
That is, bosons have $h=2$, fermions have $h=1$ and we 
can see that for any $h$ between these 
two extremes we obtain particles with different fractional 
spins with the same value of $h$, that is, {\it this number 
also classifies the nature of the fractional spin particles}.
  
Now, we consider an anyonic system of two particles with
the same fractional spin. We observe that $h$ of each 
particle is different from $h_{s}$, that is, the global 
statistics. We have pondered the 
significance of these different numbers, and conjectured 
that they can be related to the interaction 
between these particles\cite{R1}. Perhaps, the Chern-Simons 
effective potential gives us some hints. Before 
we write down an expression for this potential, some words about 
interaction are called for. Only when the particle $p_{1}$ 
passes along trajectories around the particle $p_{2}$ can we 
speak about interaction, that is, considering the Chern-Simons 
effective potential in the 
non-relativistic case as given in\cite{R4}

\begin{equation}
\label{f1}
I_{12}=2\int\; dt\frac{d\Theta(x_{1}-x_{2})}{dt}+I_{g},
\end{equation}

\noindent  where $\Theta$ is a polar angle defined by the 
relative position of the particles and $I_{g}$ is a term 
that vanishes for closed paths and is 
unimportant for open paths since the statistics and angular 
momentum do not change. We have observed that for one particle 
the Hausdorff dimension is greater 
than that for a system of two particles with the same spin. 
This means that, once the Lorentz force vanishes because 
self-interaction is absent\cite{R3},
the number $h$, alone characterizes the particle. On the other
hand, $I_{g}$ does not contribute to the potential when the 
paths are closed, that is, when we have interaction between 
the particles. So we can say that the number $h_{s}$ expresses 
this fact because it has a value less than $h$. Now, the effective 
potential for closed paths is

\begin{equation}
\label{b1}
I_{12}=2\int\;dt\;\frac{d\Theta(x_{1}-x_{2})}{dt},
\end{equation}

\noindent with the following meaning: it represents
the total angle
of rotation which is equal to $2\pi$ times their
linking number,
$m(\gamma_{1},\gamma_{2})$, that is, the number of
times the two paths,
$\gamma_{1}$ and $\gamma_{2}$, link.
  
In summary, $h_{s}< h$, because there is an interaction 
between two particles that sweep out closed paths 
with non-zero linking number. As we said at the 
beginning, the Hausdorff dimension can be expressed 
in terms of the statistics, so the following interpretation 
may be made, {\it when interaction occurs, the Hausdorff 
dimension changes because the statistics change}, or 
alternatvely, {\it the Chern-Simons field in addition to 
changing the statistics of the matter 
field also modifies the global statistics of the 
particles when they are interacting}. In this way, 
we consider the equivalence classes of the trajectories 
of these particles and may thus sketch a new model 
of anyonic interaction.

\acknowledgments
I would like to thank Steven F. Durrant for reading
 the manuscript.

\end{document}